\documentclass[twocolumn,trackchanges]{aastex63}
\RequirePackage{lineno}
\usepackage{amsmath}
\usepackage{amssymb}
\usepackage{amsthm}
\usepackage{natbib,aasdefs,url,bm}
\usepackage{array}
\usepackage{float}
\usepackage{graphicx}
\usepackage{subfigure}
\usepackage{color}
\usepackage{ctable}
\usepackage{threeparttable} 
\usepackage{lineno}
\usepackage{tablefootnote}

\def\hlinewd#1{%
\noalign{\ifnum0=`}\fi\hrule \@height #1 %
\futurelet\reserved@a\@xhline} 

\shorttitle{TeV Gamma-Rays from the LLAGN NGC 4278}
\shortauthors{Yuan and Liu}

\begin{document}

\title{TeV Gamma-Rays from the Low-Luminosity Active Galactic Nucleus NGC 4278: Implications for the Diffuse Neutrino Background}%

\correspondingauthor{Chengchao Yuan}
\email{chengchao.yuan@ulb.be}

\author[0000-0003-0327-6136]{Chengchao Yuan}\affil{Service de Physique Théorique, Université Libre de Bruxelles, CP225 Boulevard du Triomphe, 1050 Brussels, Belgium}

\author[0000-0003-1576-0961]{Ruo-Yu Liu}\affil{School of Astronomy and Space Science, Nanjing University, Nanjing 210023, China}\affil{Key Laboratory of Modern Astronomy and Astrophysics (Nanjing University), Ministry of Education, Nanjing 210023, China}


\begin{abstract}
This work investigates the origin of the TeV emission detected by the Large High Altitude Air Shower Observatory (LHAASO) from NGC~4278, a galaxy hosting a low-luminosity active galactic nucleus (LLAGN). Considering two plausible scenarios, AGN jets and winds, we model the X-ray, GeV, and TeV emission during both TeV-low (quasi-quiet) and TeV-high (active) states. The spectral energy distributions can be explained either by single-zone leptonic emission from moderately relativistic jets or by lepto-hadronic emission from sub-relativistic winds. The best-fit parameters suggest that the transition from the quasi-quiet to the active state may be driven jointly by an enhanced accretion rate and the jet deceleration or wind expansion. We further show that future MeV and very-high-energy $\gamma$-ray observations can discriminate between the {leptonic and lepto-hadronic scenarios}. Although the neutrino flux from {NGC 4278} predicted by the wind model is too low to be detected with current neutrino observatories, a lepto-hadronic wind scenario can account for the PeV diffuse neutrino background when adopting a local LLAGN density {($n_{\rm L,0}$) corrected for the TeV duty cycle ($\Delta T_{\rm TeV}/T$, the fraction of a LLAGN's lifetime spent in a TeV-emitting phase)}, $n_{\rm L,0}(\Delta T_{\rm TeV}/T) \sim 10^{-5}~\rm Mpc^{-3}$, as inferred from the LHAASO detection.

\end{abstract}


\section{Introduction}\label{sec:intro}
Active galactic nuclei (AGNs), powered by persistent accretion onto supermassive black holes (SMBHs) at the galaxy central regions, are among the most prominent source classes in the $\gamma$-ray sky. Observationally, a key feature of AGNs is their broadband emission, spanning radio, infrared, optical/ultraviolet, X-ray, and $\gamma$-ray wavelengths. A subset has also been identified as potential neutrino emitters \citep[e.g.,][]{IceCube:2018dnn,IceCube:2019cia,IceCube:2022der}, indicating the presence of high-energy particle acceleration involving both protons and electrons. Although radio-loud AGNs—particularly blazars with relativistic jets aligned toward Earth—constitute the majority of sources detected in high-energy surveys, the Large High Altitude Air Shower Observatory (LHAASO) recently reported the TeV source 1LHAASO J1219+2915 associated with the low-luminosity AGN (LLAGN) NGC 4278 \citep{LHAASO:2024qzv}, which exhibits characteristics of a low-ionization nuclear emission-line region (LINER) \citep{Giroletti:2004bw,Yuan:2009jk}. This detection was made with high statistical significance, e.g., $\sim8.8\sigma$ during its active state, and with an angular offset of only $\sim0.03$ degree.

As the most abundant class of AGNs, LLAGNs are characterized by a low ratio of bolometric to Eddington luminosity (i.e., $L_{\rm bol}/L_{\rm Edd}\sim10^{-6}-10^{-4}$) and faint radio emission \citep[e.g.,][]{Ho:2008rf}, indicative of inefficient accretion and weak jet activity. For example, NGC 4278 has an Eddington ratio of $\sim5\times10^{-6}$ \citep{2010A&A...517A..33Y,Hernandez-Garcia:2014ara}, and Very Long Baseline Array (VLBA) observations reveal a sub-relativistic, S-shaped jet extending to $\sim$ a few $\times$ pc \citep{Giroletti:2004bw}. One intriguing feature of NGC 4278 is that it is the first LLAGN with TeV $\gamma$-rays observed by LHAASO. The variable TeV light curve exhibits two distinct states: an active state and a quasi-quiet state, with the fluxes differing by a factor of $\sim7$ \citep{LHAASO:2024qzv}. In addition, dedicated analyses of \emph{Fermi} Large Area Telescope (\emph{Fermi}-LAT) data have revealed $\gamma$-ray detection above 8 GeV during the active state \citep{Bronzini:2024vll}, during which enhanced X-ray fluxes are also detected by the \emph{Swift} X-Ray Telescope (XRT), in contrast to the soft archival X-ray data. These variable features provide useful information for inferring the transition of physical conditions between the quasi-quiet and active states.

The detection of TeV $\gamma$-rays may hint at particle acceleration activities in LLAGNs. Following LHAASO's discovery, leptonic radiation by accelerated electrons in single-zone jets has been proposed as the origin of the $\gamma$-rays \citep{Dutta:2024yws,Lian:2024xnb}, in which synchrotron and inverse Compton (IC) emissions in the synchrotron self-Compton (SSC) scenario are used to reproduce the X-ray and $\gamma$-ray spectra, respectively. Moreover, a hadronic origin within an extended massive molecular cloud ($\sim$ 15 kpc) around the nucleus has also been investigated by \cite{Shoji:2025znc}. 

Building on this progress and guided by the presence of sub-relativistic winds from LLAGNs driven by radiatively inefficient accretion flows \citep[RIAFs, e.g.,][]{Narayan:1994xi,Ho:1997vg,Yuan:2014gma}, which have been proposed as a potential origin of $\gamma$-rays \citep{Mahadevan:1997qz,Niedzwiecki:2013gla,deMenezes:2020rah} and diffuse neutrinos \citep[e.g.,][]{Kimura:2014jba,Fujita:2015xva,Kimura:2020thg}, this work systematically investigates the origin of X-ray and $\gamma$-ray emission in both quasi-quiet and active states from AGN components, including single-zone jets and winds. Synchrotron and SSC emission are considered in the jet scenario to account for the X-ray and $\gamma$-ray measurements, while a lepto-hadronic framework is adopted in the wind scenario. In this context, radiation produced by secondary particles from hadronic interactions reproduces the X-ray and $\gamma$-ray observations, whereas the low-energy archival data constrains the leptonic injection. The best-fit parameters obtained from the scans are then used to trace the evolution of jets and winds, including variations in the accretion rate and the size of the emission regions. We further evaluate the contribution of LLAGN lepto-hadronic winds to the diffuse neutrino background, using the TeV duty cycle (the fraction of time NGC~4278 emits at TeV energies) inferred from LHAASO observations.

Section~\ref{sec:obs} presents a review of the multiwavelength observations of NGC~4278 and the construction of the spectral energy distributions (SEDs) for the quasi-quiet and active states. The jet and wind models, together with the resulting SED fits, are presented in Section~\ref{sec:model}. Section~\ref{sec:discussion} discusses how to distinguish between these scenarios, as well as the potential contribution of lepto-hadronic LLAGN winds to the diffuse neutrino background. The main findings are summarized in Section~\ref{sec:summary}.

\section{Multiwavelength Observations}\label{sec:obs}
NGC~4278 is a nearby ($z \ll 0.1$) elliptical galaxy located at a luminosity distance of $d_L = 16.4$~Mpc \citep{Tonry:2000aa}. Its nuclear region has been identified as a LLAGN \citep[SMBH mass $M_{\rm BH} \simeq 3.1 \times 10^8,M_\odot$, ][]{2003MNRAS.340..793W,Chiaberge:2005uy}. Prior to the recent $\gamma$-ray detections by LHAASO and \emph{Fermi}-LAT, as well as the serendipitous X-ray observations by \emph{Swift}-XRT in the active state, this source had already been observed in the radio, infrared, optical, ultraviolet (UV), and X-ray bands, constituting the archival data set. 

Here we review the multiwavelength observations and construct the SEDs in both active and quasi-quiet states following \cite{Bronzini:2024vll}. For clearance, the observations are divided into the following three categories.

\begin{figure*}[htp]\centering
    \includegraphics[width=0.49\textwidth]{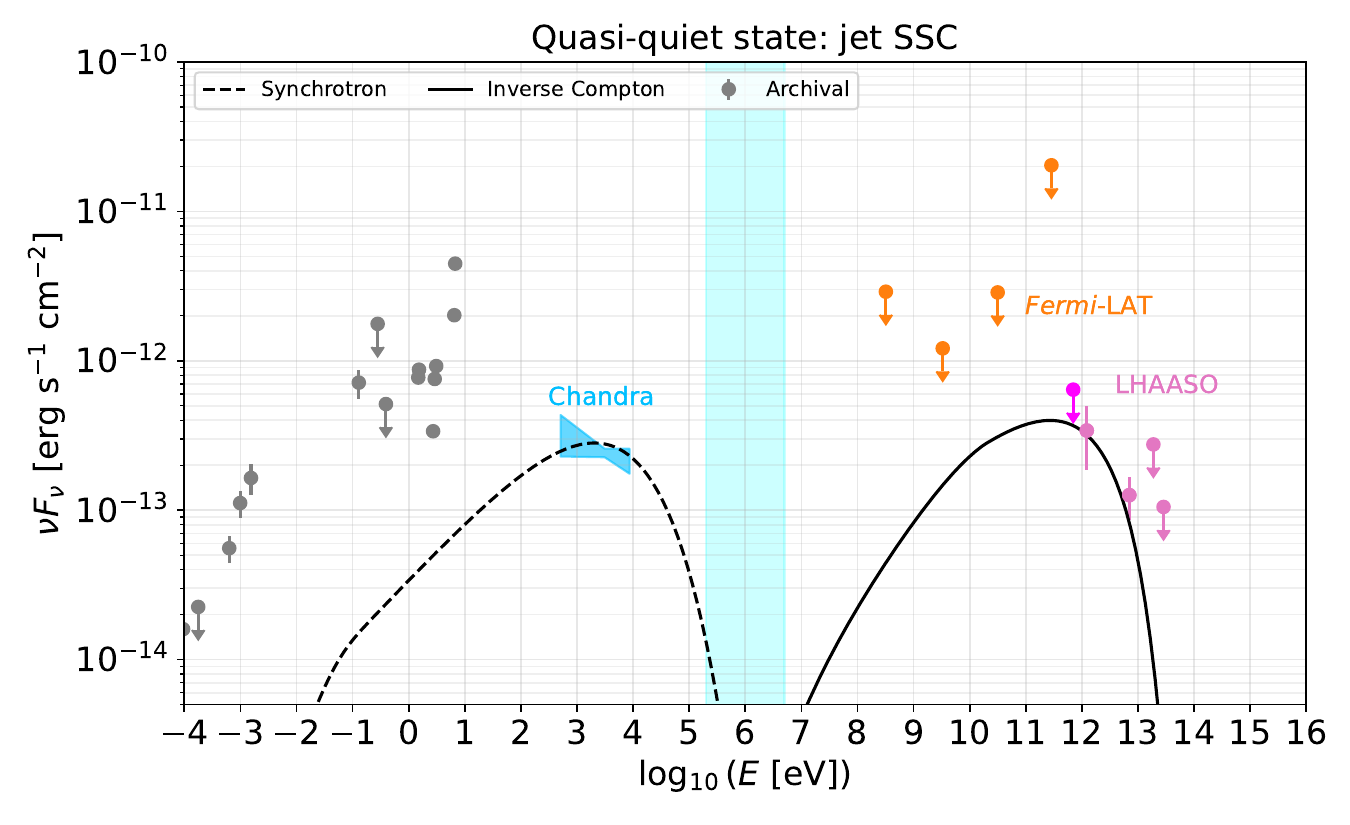}
        \includegraphics[width=0.49\textwidth]{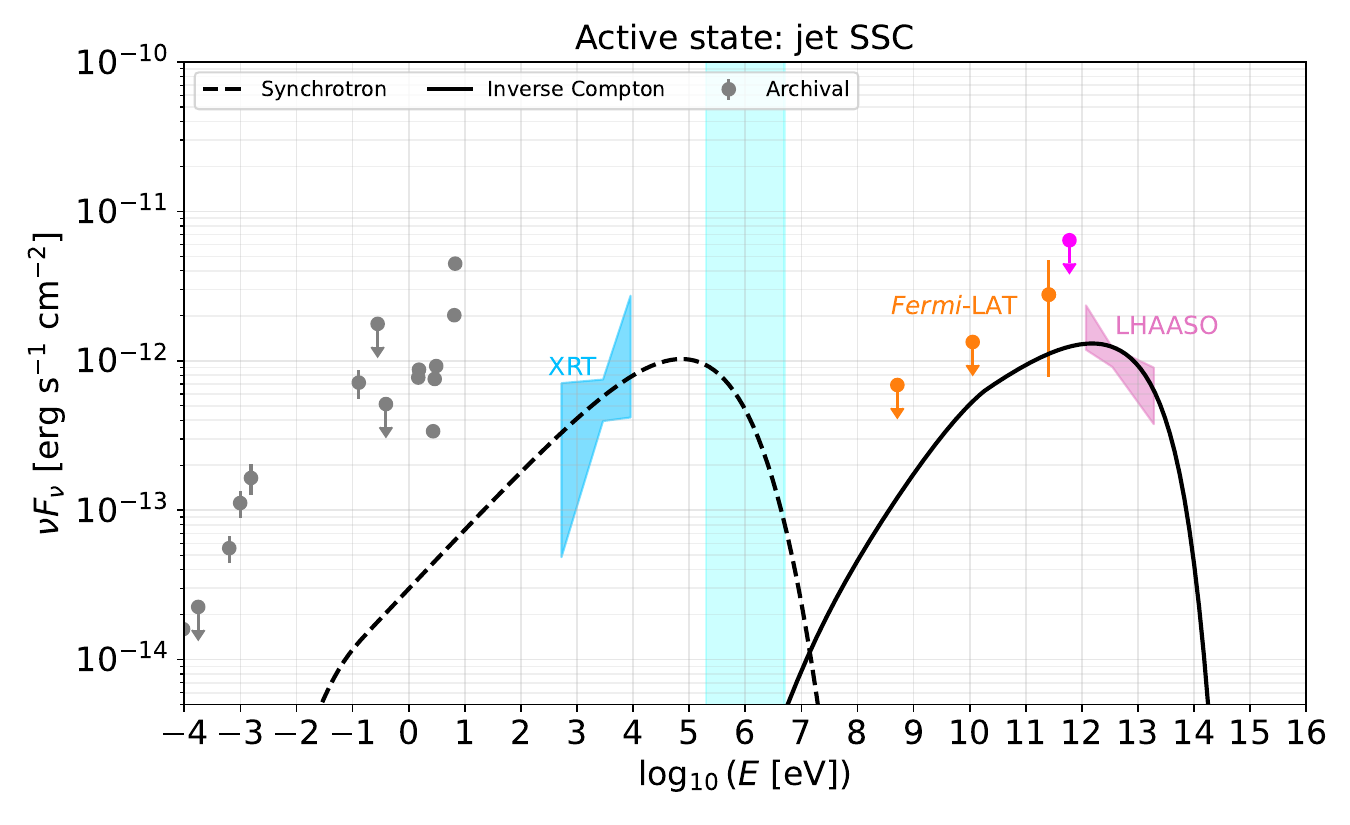}

    \caption{Jet SSC scenario: SED fitting for NGC 4278 in the quasi-quiet (left panel) and active (right panel) states. The gray, orange, magenta, and purple data points respectively represent the archival radio/infrared/optical data, the Fermi-LAT, {VERITAS}, and LHAASO measurements, while the blue regions depict the X-ray observations from \emph{Swift}-XRT. The solid and dashed curves respectively demonstrate the synchrotron and IC components. The cyan regions depict the MeV band. Data sources: \cite{LHAASO:2024qzv,Bronzini:2024vll,VERITAS:2026pcq,Anton:2004bm,Giroletti:2004bw,Cardullo:2009zw,2010A&A...517A..33Y,Pellegrini:2012ws,2021ApJ...910..104I}.}
    \label{fig:jet_SEDs}
\end{figure*}

{\it Archival radio, infrared, optical, and UV data} (gray points in Figure \ref{fig:jet_SEDs}). NGC 4278 has been extensively observed at low energies since the early 2000s, with archival data revealing a low-power radio galaxy hosting a young compact symmetric objects \citep[CSO,][]{Kiehlmann:2023shs,Readhead:2023lyp}. VLBA observations at GHz frequencies resolve a two-sided, pc-scale jet emerging from a flat-spectrum core \citep{Giroletti:2004bw}, with a radio spectrum peaking at $\sim$ 1 GHz \citep{,2016MNRAS.459..820T}, characteristic of synchrotron self-absorbed emission. From infrared to UV bands, the emission is dominated by the old stellar population and dust in the host galaxy \citep{Shapiro:2009wt,Kuntschner:2010ch}. High-angular-resolution optical and UV observations with Hubble Space Telescope (HST) reveal a faint, barely resolved nuclear component \citep{Cardullo:2009zw}, typical of LINER nuclei \citep{Maoz:2005ak}. These archival data provide constraints on the low-energy SED, which likely represents an average or quasi-steady nuclear state \citep{Nemmen:2013mya}.
    
{\it X-rays and $\gamma$-rays in the quasi-quiet state.} In the TeV quasi-quiet state, NGC 4278 has been observed with Chandra and XMM-Newton, which resolve a dominant nuclear X-ray point source embedded in diffuse thermal emission from the host galaxy \citep[see, e.g.,][and references therein]{2001ApJ...549L..51H, 2010A&A...517A..33Y,Hernandez-Garcia:2014ara,Bronzini:2024vll}. The nuclear spectrum is well described by a power law with photon index $\Gamma \simeq 2.2-2.4$, with 0.5–8 keV luminosities ranging up to a few $\times10^{40}~\rm erg~s^{-1}$ (see the blue region in the left panel of Figure \ref{fig:jet_SEDs}). At GeV energies, the source is not detected in long-term averaged Fermi-LAT catalogs, implying that any persistent GeV emission remains below current sensitivity limits \citep[orange points,][]{2022ApJS..260...53A,Ballet:2023qzs,Lian:2024xnb}. At TeV energies, however, LHAASO reports a steady low-flux component outside the main active phase, detected at a relatively lower significance and flux level than during the active phase, shown as the purple data points \citep{LHAASO:2024qzv}. This quasi-quiet TeV emission suggests ongoing particle acceleration even during periods of reduced activity, and provides an important baseline for comparison with the enhanced active state.
    
{\it X-rays and $\gamma$-rays in the active state.} A pronounced active state was observed during the LHAASO campaign between 2021 and 2022, when NGC 4278 was detected as a TeV source by LHAASO Water Cherenkov Detector Array (WCDA), with evidence for a significantly enhanced flux over several months \citep{LHAASO:2024qzv}. During the same period, a dedicated analysis of contemporaneous \emph{Fermi}-LAT data reveals a statistically significant GeV detection above $\sim$~8 GeV \citep{Bronzini:2024vll}, characterized by a hard spectrum (spectral index $\Gamma\sim1.3$) and a $\gamma$-ray luminosity of $\sim10^{41}~\rm erg~s^{-1}$. A serendipitous \emph{Swift}-XRT observation obtained in November 2021 shows the nucleus in a high X-ray state \citep{Bronzini:2024vll}, with a 0.5$-$8 keV flux comparable to the brightest levels previously observed by Chandra. The purple region, orange points and blue region in the right panel of Figure \ref{fig:jet_SEDs} respectively illustrate the LHAASO, \emph{Fermi}-LAT and \emph{Swift}-XRT observations. The temporal coincidence of enhanced X-ray, GeV, and TeV emission strongly suggests an episode of elevated activity, which is adopted as the active-state SED.


\section{SED Modeling}\label{sec:model}
\label{sec:modeling} 

This section presents two possible scenarios for SED fitting: relativistic jets and sub-relativistic winds. 

\subsection{Jet scenario}\label{sec:model_jet}

We consider a compact radiation region of plasma containing relativistic electrons that moves along the jet axis as a spherical ``blob" of radius $R_b'$, permeated by a magnetic field of strength $B_b'$, and propagating with a bulk Lorentz factor $\Gamma_b$. Here and hereafter, the primed symbols are used to denote the quantities measured in the rest frame of the relativistic jet (e.g., the comoving frame). We adopt a power-law spectrum with the spectral index $s$ and an exponetial cutoff to describe the proton injection rate as a function of electron Lorentz factor ($\gamma_e'$), e.g., $Q_{e}'\propto\gamma_e'^{-s}\exp(-\gamma_{e}'/\gamma_{e,\rm max}')$ for $\gamma_e'>\gamma_{e,\rm min}'$, where $\gamma_{e,\rm min}'$ and $\gamma_{e,\rm max}'$ represents the minimum and maximum electron Lorentz factors. Given the injection luminosity $L_e'$, the injection rate could be normalized via $(4\pi R_b'^3/3)\int_{\gamma_{e,\rm min}'}^\infty\gamma_e'Q_{e}'d\gamma_e'=L_e'/(m_ec^2)$. 

In the radiation zone, the injected electrons lose energy through synchrotron and SSC cooling processes and can escape on a timescale $t_{\rm esc}'$. The electron distribution can be described by a time-dependent transport equation,
\begin{linenomath}
\begin{equation}
    \frac{\partial n_e'}{\partial t'}=\dot Q_{e}'-\frac{\partial}{\partial \gamma_e'}\left(\dot \gamma_e'n_e'\right)-\frac{n_e'}{t_{\rm esc}'},
    \label{eq:transport}
\end{equation}
\end{linenomath}
where $t'$ denotes the time measured in the comoving frame, $n_e'\equiv d^2N_e'/(d\ln\gamma_e'dV')$ is the electron number density, and $\dot \gamma_e'\equiv \gamma_e'(t_{\rm sy}'^{-1}+t_{\rm ssc}'^{-1})$ represents the cooling rate due to synchrotron ($t_{\rm sy}'^{-1}$) and SSC ($t_{\rm ssc}'^{-1}$) radiation. In this model, the electron distribution can eventually reach a steady state in which electron losses are balanced by continuous injection. Assuming that electrons escape on the light-crossing timescale, $t_{\rm esc}' = R_b'/c$, and that the jet is oriented toward the Earth, the open-source software AM$^3$ \citep[Astrophysical Multi-Messenger Modeling][]{Klinger:2023zzv} is used to solve the time-dependent electron and photon distributions until a steady state is reached, from which the observed photon flux (in the units of $\rm erg~s^{-1}~cm^{-2}$) can be written as
\begin{linenomath}
\begin{equation}
    \nu F_\nu(E_\gamma)=\frac{\Gamma_b^3}{3}\left(\frac{R_b'^2}{d_L^2}\right)c\varepsilon_\gamma'n'_\gamma(\varepsilon_\gamma'),
\end{equation}
where $n_\gamma'$ is the comoving photon density and the observed photon energy ($E_\gamma$) is related to the comoving photon energy ($\varepsilon_\gamma'$) via $E_\gamma=\Gamma_b\varepsilon_{\gamma}'$.\end{linenomath}

Our primary goal is to fit the X-ray and $\gamma$-ray spectra using a relativistic jet likely emerging just before or during the TeV-emitting epoch, since the low-energy archival emission originates from non-AGN components (e.g., stellar light and dust) or from average nuclear activity, such as the accretion disk or weak outflows. The $\chi^2$ quantities for the quasi-quiet and active states is employed to evaluate the goodness of fit and to determine the best-fit parameter sets. To simplify the parameter scan and to capture the time evolution of the jet parameters, we fix $\gamma_{e,\rm min}' = 10^4$ and the spectral index $s = 2.2$ as fiducial values\footnote{$\gamma_{e,\rm min}'$ is constrained so that the SSC spectrum peaks at $\sim \gamma_{e,\rm min}'^{2} \times (1-10)~{\rm keV} \sim 0.1-1$~TeV.}, following the results from previous studies \citep[e.g.,][]{Dutta:2024yws,Lian:2024xnb}, and allow $R_b'$, $\Gamma_b$, $B_b'$, $L_e'$, and $\gamma_{e,\rm max}'$ to vary freely. Details of the $\chi^2$ definition and the ranges of these parameters are provided in Appendix~\ref{app:fitting}. 

The best-fit parameters are listed in the upper part of Table~\ref{tab:parameters}, while the corresponding SED fits are shown in the left and right panels of Figure~\ref{fig:jet_SEDs} for the quasi-quiet and active states, respectively. The $1\sigma$ uncertainties, along with the posterior parameter distributions, are presented in Appendix \ref{app:fitting} (Figure~\ref{fig:corner_plots}). 
The fitting indicates that a moderately relativistic jet with Lorentz factors of $\Gamma_b \simeq 3-4$ is favored. Meanwhile, the transition trend between the quasi-quiet and active states, namely (1) increasing $R_b'$ and $L_e'$ and (2) decreasing $\Gamma_b$ and $B_b'$, suggests a decelerating jet blob with enhanced particle injection. The increasing maximum electron Lorentz factor $\gamma_{e,\rm max}'$ can also be qualitatively interpreted as a consequence of a weakening magnetic field, since it can be estimated by equating the synchrotron cooling time $t_c' = 6\pi m_e c / (\gamma_e' \sigma_T B_b'^2)$ to the acceleration time $t_{\rm acc}' \propto \gamma_e' m_e c / (e B_b')$, where $\sigma_T$ is the Thomson cross section. {To justify how the fitting depends on the fixed parameters, we computed the SSC spectra by varying $s$ and $\gamma'_{e,\rm min}$ in the ranges $[2.0,~2.4]$ and $[10^{3},~10^5]$. The results demonstrate that the best-fit parameters in Table~\ref{tab:parameters} do not change significantly when $s$ and $\gamma'_{e,\rm min}$ vary within the ranges $2.0 \lesssim s \lesssim 2.3$ and $10^{3.6} \lesssim \gamma'_{e,\rm min} \lesssim 10^4$.}

\begin{table}
	\caption{Best-fit parameters for the jet and wind scenarios in the quasi-quiet (Q.-Q.) and active states. }
	\label{tab:parameters}

       {\bf Jet leptonic scenario ($\gamma_{e,\rm min}'=10^4$, $s=2.2$)}
    \\
    \begin{tabular}{lccccc} 
		\hline
        \hline
		  & $R'_b$ & $\Gamma_b$ & $B_{b}'$ & $L_e'$ & {$\gamma_{e,\rm max}'$}\\
         & [cm] & - & [mG] & [$\rm erg~s^{-1}$] & -  \\
		\hline
		Q.-Q.& $6.0\times10^{15}$ & 4.1 & $4.7$ & $1.0\times10^{41}$ & $4.1\times10^6$\\
        Active & $6.9\times10^{15}$ & 3.2 & $3.1$ & $3.8\times10^{41}$ & $1.9\times10^7$\\
		\hline
	\end{tabular}

\vspace{1em} 

 {\bf Wind lepto-hadronic scenario ($s=2.0$)}\\   
	\begin{tabular}{lccccc} 
		\hline
        \hline
		{} & $R_{\rm diss}$  & $L_p$ & $E_{p,\rm max}$ & $B_{\rm diss}$& $L_e~(\lesssim)$\\
        {} &[cm]  & $\rm [erg~s^{-1}]$ & [PeV] & [G] & {$[\rm erg~s^{-1}]$}\\
		\hline
		Q.-Q. & $1.7\times10^{16}$  & $1.0\times10^{44}$ & 470 & 0.32 & $4.7\times10^{40}$\\
		Active & $7.7\times10^{16}$  & $4.6\times10^{44}$ & 190 & 0.63 & $3.3\times10^{40}$\\
		\hline
	\end{tabular}

\end{table}

\subsection{Sub-relativistic wind scenario}\label{sec:model_wind}

\begin{figure*}[htp]\centering
    \includegraphics[width=0.49\textwidth]{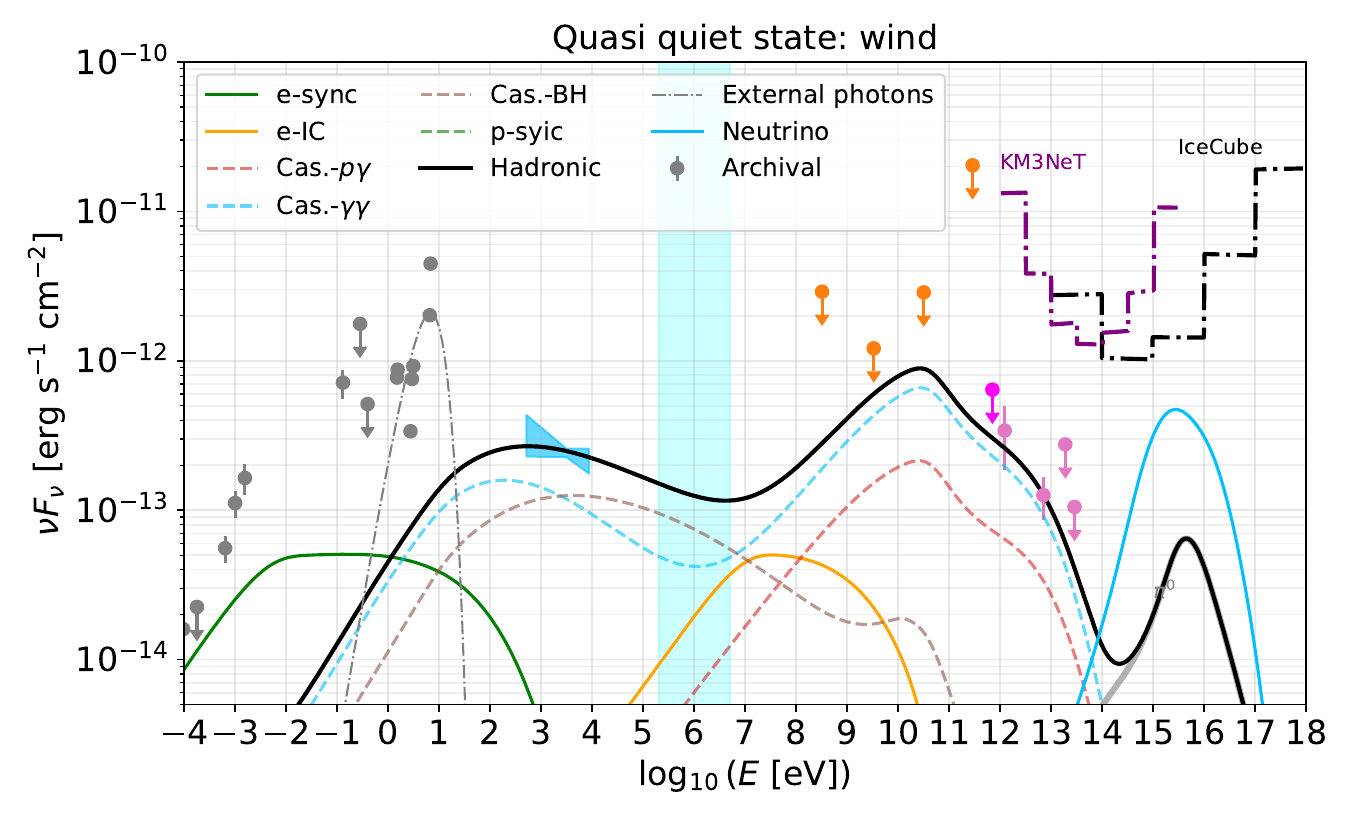}
       \includegraphics[width=0.49\textwidth]{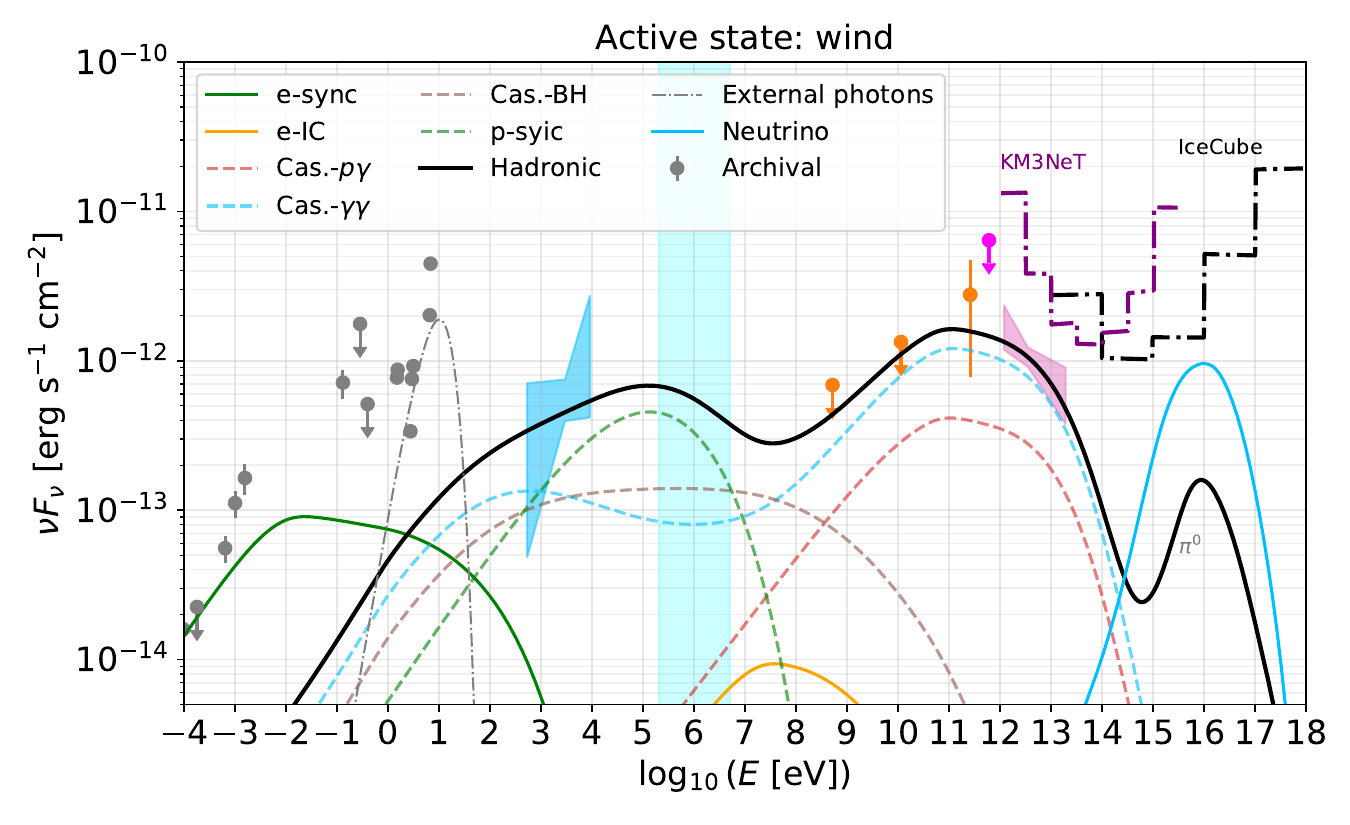}
    \caption{Wind lepto-hadronic scenario: SED fitting for NGC 4278 in the quasi-quiet (left panel) and active (right panel) states. The data points have the same meaning with these in Figure \ref{fig:jet_SEDs}. In both panels, the cascade photon spectra, the single-flavor neutrino flux, and the external photon spectra are shown as the solid black, solid blue, and dash-dotted gray curves, respectively. Contributions from different interaction components are also displayed. The green and orange solid curves represent the synchrotron and IC radiation from primary electrons. The 10-year flux sensitivities of IceCube \citep[for declination $\delta=-23^\circ$;][]{IceCube-Gen2:2021rkf} and KM3NeT/ARCA230 \citep[for $\delta=-73^\circ$;][]{KM3NeT:2024uhg}, at the 90\% confidence level, are indicated by the black and purple dash-dotted curves, respectively.}
    \label{fig:wind_SEDs}
\end{figure*}


As demonstrated in Section~\ref{sec:model_jet}, a moderately relativistic jet with a Lorentz factor of $\sim 3-4$ is favored for pure leptonic radiation; therefore, proton injection, is typically required in the sub-relativistic winds.

Power-law distributions are assumed for both electron and proton injection, $Q_{e(p)} \propto \gamma_{e(p)}^{-s} \exp(-\gamma_{e(p)}/\gamma_{e(p),\rm max})$, where $\gamma_p = E_p / (m_p c^2)$ is the proton Lorentz factor defined by the proton energy $E_p$. We model the isotropic radiation from the wind, characterized by the energy dissipation radius $R_{\rm diss}$ and magnetic field strength $B_{\rm diss}$, and use the electron (proton) luminosity $L_e$ ($L_p$) to normalize the injection rate. We fix the minimum proton energy to its rest-mass energy, $E_{p,\rm min} = m_p c^2$, and treat the maximum proton energy, $E_{p,\rm max} = \gamma_{p,\rm max} m_p c^2$, as a free parameter. In this scenario, the electron luminosity is weakly constrained, since synchrotron emission from injected electrons typically peaks at low energies, such as the radio to optical bands, whereas electromagnetic cascades induced by secondary particles produced in $p\gamma$ (photomeson) interactions extend from the X-ray to the $\gamma$-ray bands. In particular, we obtain only upper limits on $L_e$, which are set jointly by (1) the upper limits from archival radio observations and (2) constraints from TeV $\gamma$-ray measurements, since a higher synchrotron peak would render the wind optically thick to TeV $\gamma$-rays due to $\gamma\gamma$ attenuation\footnote{{Because the $\gamma\gamma$ attenuation threshold scales inversely with target photon energy, $\sim 2.2$ eV external photons dominate absorption at $\sim 100$ GeV, while lower-energy photons from injected electrons can strongly attenuate TeV emission.}}. To make the parameter scan computationally feasible (noting that lepto-hadronic simulations until steady state can be time-consuming), the parameters $\gamma_{e,\rm min} = m_p/m_e \sim 10^3$, $\gamma_{e,\rm max} = 10^6$, and $s = 2$ are fixed. 

The parameter scan scheme is outlined as follows: we first vary $R_{\rm diss}$, $L_p$, $B_{\rm diss}$, and $E_{p,\rm max}$ freely to fit the X-ray and $\gamma$-ray measurements; we then gradually increase $L_e$ from $10^{40}~\rm erg~s^{-1}$ to $10^{42}~\rm erg~s^{-1}$ until either the radio constraints or the requirement that the wind remains optically thin to TeV photons is no longer satisfied, thereby determining its upper limit. The range of each parameter can be found in Appendix \ref{app:fitting} (Table \ref{tab:ranges}).

Noting that the nuclear region, such as the LINER phenomenology, may also contribute to the archival optical and UV emission \citep{Maoz:2005ak,Cardullo:2009zw}, we treat them as external photon fields that participate in $p\gamma$ interactions. {The external optical/UV (OUV) luminosity $L_{\rm ext}$ is a key parameter, as it directly affects the $p\gamma$ efficiency, $f_{p\gamma} \propto L_{\rm ext} R_{\rm diss}^{-2}$. To estimate $L_{\rm ext}$, we use nuclear-only OUV measurements of NGC 4278 from \cite{Cardullo:2009zw}, which report fluxes of $(6.9-10.8)\times10^{-16}~{\rm erg~s^{-1}~cm^{-2}~\AA^{-1}}$, corresponding to a bolometric luminosity $L_{\rm ext} \sim (1.2-1.9)\times10^{41}~\rm erg/s$. On average, we adopt $L_{\rm ext} \simeq 1.6 \times 10^{41}~\rm erg/s$ and $k_B T_{\rm ext} \sim 2.2~\rm eV$, consistent with the measurement bandpass and yielding an OUV flux comparable to the archival data.} The external photon spectrum is shown as the gray dashed curve in Fig.~\ref{fig:wind_SEDs}.

The external photon energy density can be expressed as $u_{\rm ext} = {L_{\rm ext}}/({4\pi R_{\rm diss}^2 c})$, assuming that the optical/UV photons originate from the inner flows or accretion disk embedded within the wind, e.g., at a radius $\lesssim 10^{16}$~cm \citep[e.g.,][]{Winter:2022fpf,Yuan:2023cmd}. Moreover, $pp$ (hadronuclear) process can also produce electromagnetic cascades, but they are typically subdominant compared to $p\gamma$, as in the wind case the ratio of $p\gamma$ to $pp$ interaction rates is high \citep[e.g.,][]{Yuan:2025zwe},\begin{linenomath}
\begin{equation}
\frac{t_{p\gamma}^{-1}}{t_{pp}^{-1}} \sim 3 \left(\frac{\sigma_{p\gamma}}{\sigma_{pp}}\right) \left(\frac{v_w}{c}\right) \left(\frac{m_p c^2}{k_B T_{\rm ext}}\right) \left(\frac{\eta_{\rm rad} L_{\rm ext}}{\eta_w \dot m L_{\rm Edd}}\right) \gtrsim 10^3,
\end{equation}\end{linenomath}
where $\sigma_{p\gamma} \simeq 0.5~\rm mb$ is the $p\gamma$ cross section at the $\Delta$-resonance, $\sigma_{pp} \simeq 40~\rm mb$ is the $pp$ cross section \citep{Kafexhiu:2014cua}, $v_w \sim 0.1~c$ is the wind velocity, $\eta_{\rm rad} \sim 0.1$ is the radiation efficiency, $\eta_w \lesssim 0.1$ is the fraction of accreted mass converted into the wind \citep{2009PASJ...61L...7O,2019ApJ...880...67J}, $\dot m \sim 0.01$ is the ratio of the accretion rate to the Eddington accretion rate \citep[comparable to the values used in ][]{Kimura:2014jba,Kimura:2020thg}, and $L_{\rm Edd} \sim 3.8 \times 10^{46}~\rm erg~s^{-1}$ is the Eddington luminosity. 

For each parameter set, we use the AM$^3$ software \citep{Klinger:2023zzv} to model lepto-hadronic interactions, including the distributions of primary protons and electrons, secondary particles such as pions, muons, electrons, neutrinos (and their anti-particles), and photons produced via $p\gamma$ interactions (`Cas.-$p\gamma$'), Bethe-Heitler pair production (`Cas.-BH'), $\gamma\gamma$ attenuation and pair production (`Cas.-$\gamma\gamma$'), synchrotron/IC radiation, proton synchrotron/IC radiation (`p-syic'), and particle decays. Specifically, we solve the transport equations (similar to Equation~\ref{eq:transport}, but for distributions and quantities defined in the SMBH rest frame) for all particle species in a time-dependent manner until a steady state is reached. Instead of assuming a free-escape timescale, an energy-dependent diffusive escape time, $t_{\rm esc} = \max[R_{\rm diss}/c,~t_{\rm diff}]$ is adopted, where $t_{\rm diff} \sim R_{\rm diss}^2 / D$ is the diffusion time and $D \sim E_p c / (e B_{\rm diss})$ is the diffusion coefficient in the Bohm limit. {In contrast to the jet scenario, a diffusive escape timescale is adopted here, as stronger confinement by larger spatial scales and magnetic fields leads to more efficient particle trapping, whereas in jet blobs the escape is well approximated by the light-crossing time.}

The best-fit parameters, obtained by minimizing $\chi^2$, are listed in the lower part of Table~\ref{tab:parameters}. The posterior distributions of key parameters ($R_{\rm diss}$, $L_p$, $B_{\rm diss}$, and $E_{p,\rm max}$) are presented in Appendix \ref{app:fitting}. The left and right panels of Figure~\ref{fig:wind_SEDs} present the SED fits for the quasi-quiet and active states, respectively. The black solid curves illustrate the total cascade emission from hadronic interactions, with the individual components shown as colored dashed curves, while the synchrotron and IC emission from primary electrons are displayed as green and orange solid curves. The high-energy peak in the 100~TeV$-$1~PeV range represents $\gamma$-ray emission produced by $\pi^0$ decay, taking into account in-source $\gamma\gamma$ attenuation.

Figure \ref{fig:wind_SEDs} demonstrates that the X-ray and $\gamma$-ray observations up to $\sim10$ TeV could be described by the wind lepto-hadronic scenario as well, with the energy dissipation occurs at the radius $\sim10^{16}-10^{\rm 17}$ cm permeated by the magnetic field of strength $\sim0.3-0.6$ G. The accelerated proton luminosities of $L_p\simeq10^{44}~\rm erg~s^{-1}~cm^{-2}$ and $4.6\times10^{44}~\rm erg~s^{-1}~cm^{-2}$ are needed for the quiet and active states, corresponding to the acceleration efficiently \begin{linenomath}
\begin{equation}
    \eta_p=\frac{L_p}{\dot M_{\rm BH}c^2}\sim0.026~\text{(quasi-quiet)}-0.12~\text{(active)},
\end{equation}
\end{linenomath}
where $\dot M_{\rm BH}c^2\sim\dot m L_{\rm Edd}/\eta_{\rm rad}\sim3.8\times10^{45}~\rm erg~s^{-1}$ is the accretion power onto the SMBH with $\dot m=0.01$ and $\eta_{\rm rad}\sim 0.1$. {In this scenario, protons may be accelerated by the wind or by accretion flows near the SMBH. The wind’s proton luminosity is limited by its kinetic power, $L_p < L_w = \eta_w \dot M_{\rm BH} v_w^2 / 2 \lesssim 10^{44}~\rm erg/s$, assuming $\dot m = 0.01$, the fraction of accreted mass converted to winds $\eta_w \lesssim 0.1$, and the wind velocity $v_w/c \sim 0.1-1$, making the wind alone insufficient to reach the TeV-active luminosity ($L_p \simeq 4.6 \times 10^{44}~\rm erg/s$). RIAFs can also accelerate protons \citep[e.g.,][]{Kimura:2014jba}, with $L_p = \epsilon_{\rm cr} \dot M_{\rm BH} c^2 \approx 3.8 \times 10^{44}~\rm erg/s~(\epsilon_{\rm cr}/0.1)$, where $\epsilon_{\rm cr} \lesssim 0.1$ ensures non-thermal particles do not significantly alter the accretion flow \citep{Kimura:2014rja,Kimura:2014jba}. During the active state, the LLAGN may undergo an enhanced accretion episode with a higher $\dot m$, which, together with RIAF acceleration and possible wind re-energization, could alleviate the high proton luminosity required to explain the observed TeV-high emission.} Moreover, the electron power is constrained to be $\lesssim 3-5 \times 10^{40}~\rm erg~s^{-1}$ to avoid overshooting the radio observations and excessively attenuating the TeV emission. 

In the wind scenario, the associated neutrino spectra (see the solid blue curves) indicate that NGC~4278 can produce PeV neutrinos, but it is very unlikely to be detected as a point source owing to its inefficient accretion and radiative output. {Even in the active state, the predicted neutrino flux remains below the 10-year sensitivity limits of IceCube and KM3NeT/ARCA230 (see the black and purple dash-dotted curves). It may be marginally detectable with future $\sim10~\rm km^3$ detectors such as IceCube-Gen2 \citep{IceCube-Gen2:2020qha} and TRIDENT \citep{2023NatAs...7.1497Y}, while detection prospects are more promising for a $\sim30~\rm km^3$ facility such as the proposed High-energy Underwater Neutrino Telescope \citep[HUNT;][]{Huang:2025auf}.}. 

\section{Discussion}\label{sec:discussion}
This work has demonstrated that the TeV $\gamma$-rays from NGC 4278 together with the X-rays and GeV $\gamma$-rays in the quasi-quiet and active states can be reproduced either by the leptonic radiation from a moderately relativistic jet or by the lepto-hadronic radiation from a sub-relativistic wind. While both scenarios suggest an enhanced accretion or injection rate as this LLAGN transitions from radiation-low to radiation-high states, distinct physical conditions and parameters emerge from the scans. For instance:
\begin{itemize}
\item The jet scenario favors the launch of a relativistic jet with a Lorentz factor $\Gamma_b \sim 3-5$, where synchrotron and SSC radiation from a power-law electron injection is sufficient to reproduce the SEDs.
\item The wind scenario typically requires a proton acceleration efficiency of $\eta_{p}\sim 2.6-12\%$, within a sub-relativistic radiation zone extending to $\sim \text{a few} \times 10^{16}$~cm. Meanwhile, the maximum proton energy can reach 100~PeV to EeV, as allowed by confinement and cooling constraints, $E_{p,\rm max} \sim e B_{\rm diss} R_{\rm diss} \sim 900~{\rm PeV} (B_{\rm diss}/0.3~{\rm G}) (R_{\rm diss}/10^{16}~\rm cm)$, making this scenario relevant for PeV neutrino production.
\end{itemize}

Hence, it is of particular interest to discuss how these two scenarios can be distinguished through multiwavelength observations, and how LLAGN lepto-hadronic winds as a population contribute to the diffuse neutrino intensity.

\subsection{Distinguishing the leptonic and lepto-hadronic origins}
Comparing the SED fits in Figure~\ref{fig:jet_SEDs} with those in Figure~\ref{fig:wind_SEDs}, a key difference between the jet SSC and wind lepto-hadronic scenarios is seen in the flux level and spectral shape in the MeV range (the cyan regions in these figures). The physical interpretation is that a higher minimum elecron Lorentz factor need to reach TeV would create a giant dip between the keV synchrotron hump and the TeV SSC hump with the energy difference by a factor of $E_{\gamma,\rm SSC}/E_{\gamma,\rm sync}\sim\gamma_{e,\rm min}'^2=10^8$, which naturally predicts a faint and soft MeV spectrum. On the other hand, multiple interaction channels in the hadronic processes can produce secondary particles across a wide range of energies, which reprocess energy into the radiation fields and consequently lead to a more continuous electromagnetic spectrum with a flatter shape. Observations by current \citep[e.g., SVOM,][]{Wei:2016eox}, near-term \citep[e.g., COSI,][]{Tomsick:2019wvo}, and planned (e.g., e-ASTROGAM \citep{e-ASTROGAM:2016bph,e-ASTROGAM:2017pxr} and AMEGO-X \citep{Caputo:2022xpx}) missions covering the MeV band ($\sim0.1-10$ MeV) can provide crucial diagnostics for distinguishing leptonic from lepto-hadronic emission, either through real-time follow-ups or multi-year surveys.

Moreover, in the GeV band, the wind scenario predicts higher $\gamma$-ray fluxes (by a factor of $\sim2-3$), dominated by $e^{\pm}$ pairs produced via efficient $\gamma\gamma$ attenuation, with a spectral peak at $\sim10-100$ GeV. This emission is more stringently constrained by the \emph{Fermi}-LAT upper limits (e.g., the black curve in the right panel of Figure~\ref{fig:wind_SEDs}). Together with the $\pi^{0}$-decay component extending to PeV energies, these features imply that the lepto-hadronic scenario in LLAGNs can be more effectively tested through broadband $\gamma$-ray observations spanning from $\sim0.1$ GeV to PeV energies, using \emph{Fermi}-LAT \citep{Fermi-LAT:2009ihh}, the Cherenkov Telescope Array \citep[CTA,][]{CTAConsortium:2010umy}, and the proposed Large Array of Imaging Atmospheric Cherenkov Telescope \citep[LACT,][]{Zhang:2024ztw}. 

{One caveat is that these model-selection criteria are primarily effective in distinguishing leptonic from lepto-hadronic emission mechanisms, but are less decisive in separating jet and wind scenarios, since jets can also generate cascade emission through proton injection. In particular, a soft and weak MeV signal favors a purely leptonic jet origin, whereas a harder MeV spectrum can arise from hadronic processes in either jets or winds.}

\subsection{LLAGN wind contribution to the diffuse neutrino intensity}

Another signature of the wind scenario is the production of PeV neutrinos. Although NGC 4278 itself is unlikely to be detected by IceCube, evaluating the contribution of the LLAGN population to the diffuse neutrino intensity during their TeV-emitting duty cycle remains valuable.

The neutrino spectra shown in Figure~\ref{fig:wind_SEDs} are used to obtain the (single-flavor) neutrino luminosities of NGC~4278-like LLAGNs in both the quasi-quiet and active states, denoted as $L_{\nu}^{(Q)}$ and $L_{\nu}^{(A)}$, respectively. The cumulative neutrino intensity as a function of neutrino energy $E_\nu$ can then be computed by integrating over redshift from $0$ to $z_{\rm max}$\footnote{Adopting $z_{\rm max}\sim5$ is sufficient to ensure convergence of the integral.}, which can be written explicitly as \citep[e.g.,][]{Alvarez-Muniz:2004xlu,Murase:2014foa}\begin{linenomath}
\begin{equation}
    \Phi_\nu(E_\nu)=\frac{c}{4\pi H_0}\int_0^{z_{\rm max}} dz \frac{n_{\rm L}(z)(\Delta T_{\rm TeV}/{T})}{\sqrt{\Omega_m(1+z)^3+\Omega_\Lambda}}\frac{\tilde L_\nu(E_\nu')}{E_\nu'^2},
\end{equation}\end{linenomath}
where $H_0 \sim 70~\rm km~s^{-1}~Mpc^{-1}$ is the Hubble constant, $n_{\rm L}(z)$ is the redshift evolution of the LLAGN number density (in units of $\rm Mpc^{-3}$), $\Delta T_{\rm TeV}/T$ denotes the TeV duty cycle (the fraction of time that an LLAGN emits at TeV energies), $\tilde L_\nu = \zeta L_\nu^{(A)} + (1-\zeta)L_\nu^{(Q)}$ is the neutrino luminosity averaged over the active and quasi-quiet states with the active-state weighting factor $0 \le \zeta \le 1$, and $E_\nu' = (1+z)E_\nu$. The LLAGN density can be generally expressed as $n_{\rm L}(z)=n_{\rm L,0}\phi(z)$, where $n_{\rm L,0}$ is the local density and $\phi(z)$ describes the redshift evolution. We adopt the redshift distribution of BL Lac objects \citep[e.g.,][]{Ajello:2013lka}, given that LLAGNs and BL Lac objects both exhibit faint accretion disk components.

\begin{figure}
    \centering
    \includegraphics[width=1.0\linewidth]{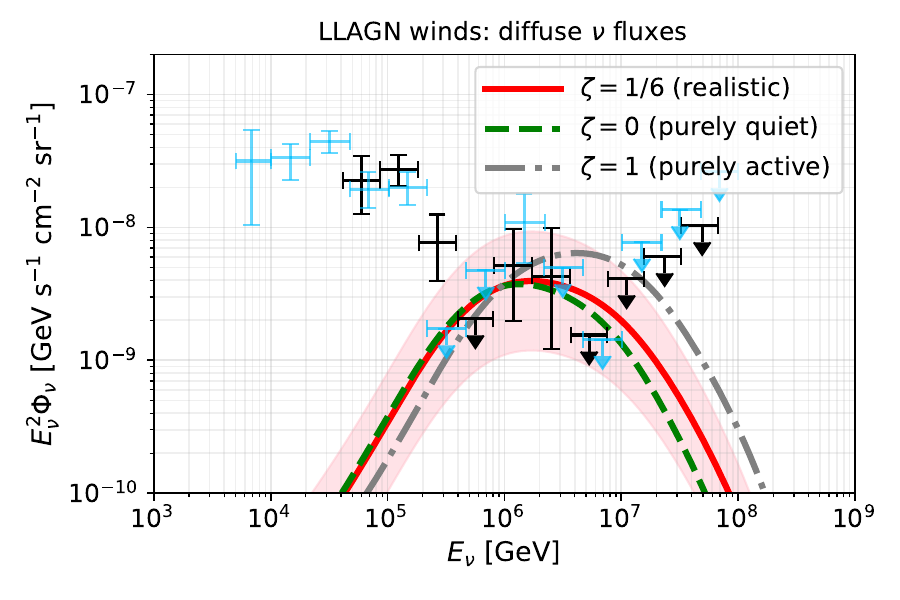}
    \caption{The cumulative diffuse neutrino spectra (single-flavor) from LLAGN winds that emit at TeV energies. The black and blue points represent the diffuse neutrino fluxes inferred from IceCube high-energy starting events analysis \citep[][]{IceCube:2024fxo} and six-year shower analysis \citep{IceCube:2020acn}, respectively. The wind model predictions for $\zeta = 1/6$, $\zeta = 0$, and $\zeta = 1$ are respectively shown as the solid red, dashed green, and dash-dotted curves. {{The red area illustrates the $1\sigma$ uncertainty for $\zeta=1/6$ case.}}}
    \label{fig:diff_neu}
\end{figure}

One could roughly estimate $n_{\rm L,0}(\Delta T_{\rm TeV}/T)$ which represents the density of local LLAGN undergoes TeV-emitting phase, as \begin{linenomath}
\begin{equation}
   {n_{\rm L,0}\left(\frac{\Delta T_{\rm TeV}}{T}\right)\sim\frac{N_{\rm L, TeV}}{4\pi D_{\rm H}^3/3}\sim N_{\rm L, TeV}\times10^{-5}~\rm Mpc^{-3}}
   \label{eq:density}
\end{equation}\end{linenomath}
where $D_{\rm H} \sim d_L \sqrt{F_{\rm obs}({\rm TeV})/F_{\rm lim}({\rm TeV})}$ is the LHAASO detection horizon for a NGC~4278-like LLAGN, $F_{\rm obs}({\rm TeV})$ is the observed TeV flux from NGC~4278, $F_{\rm lim}({\rm TeV})$ is the LHAASO-WCDA sensitivity for a $5\sigma$ detection over a 1.6 yr exposure \citep{LHAASO:2019qtb}, and {$N_{\rm L, TeV}$ is the number of LLAGNs that emit TeV $\gamma$-rays. Since NGC 4278 is the only LLAGN reported by LHAASO during its 1.6-year campaign between March 2021 and October 2022, we adopt the Poission likelihood $\mathcal L(N_{\rm L,TeV})\propto N_{\rm L,TeV} e^{-N_{\rm L,TeV}}$, with $N_{\rm L,TeV}=1$ as the most probable value. From this distribution, we infer $0.3\lesssim N_{\rm L,TeV}\lesssim2.4$ as its $1\sigma$ uncertainty. Equation \ref{eq:density} further implies a TeV duty cycle of $\Delta T_{\rm TeV}/T \sim 10^{-3}$, indicating that only about $0.1-1\%$ of an LLAGN's lifetime is spent in a TeV-emitting phase, given the local LLAGN number density of $\sim10^{-3}-10^{-2}~\rm Mpc^{-3}$ \citep[e.g.,][]{Ho:2008rf} from optical and/or X-ray observations.}

Assuming a $\Lambda$CDM universe with $\Omega_m=0.27$ and $\Omega_\Lambda=1-\Omega_m$, Figure~\ref{fig:diff_neu} presents the diffuse neutrino flux for three cases for $N_{\rm L,TeV}=1$: (1) $\zeta=1/6$ (red solid curve), corresponding to a realistic situation in which the active state lasts for approximately one-sixth of the 1.6 yr LHAASO observation period \citep{LHAASO:2024qzv}; (2) $\zeta=0$ (green dashed curve), where only the quasi-quiet state contributes; and (3) $\zeta=1$ (gray dashed-dotted curve), an optimistic case in which the entire TeV-emitting phase occurs at the active state. {The red area indicates the $1\sigma$ uncertainty for $\zeta=1/6$ case.} These results indicate that the LLAGN lepto-hadronic wind scenario with $\zeta=1/6$ and $\zeta=0$ can account for a significant fraction {($\sim30\%-100\%$)} of the PeV diffuse neutrino intensity, whereas the optimistic case with $\zeta=1$ is ruled out by upper limits at $\sim10$ PeV. Given $\tilde L_\nu\simeq10^{40.4}~\rm erg~s^{-1}$, the red curve in Figure \ref{fig:diff_neu} is consistent with the theoretical estimation for single-flavor neutrinos \citep[e.g.,][]{Murase:2014foa}\begin{linenomath}
\begin{equation}\begin{split}
    E_\nu^2\Phi_\nu&\approx\frac{c}{4\pi H_0}\xi_zn_{\rm L,0}\left(\frac{\Delta T_{\rm TeV}}{T}\right)\tilde L_{\nu}\\
    &\simeq3\times10^{-9}{~\rm GeV~s^{-1}~cm^{-2}~sr^{-1}},
    \end{split}
\end{equation}\end{linenomath}
where $\xi_z\sim0.5$ is the redshift correction factor \citep{Murase:2016gly,Yuan:2019ucv} for sources with the same redshift distribution as BL Lac objects.

\section{Summary and conclusions}\label{sec:summary}
A systematic investigation of the jet leptonic and wind lepto-hadronic scenarios for the LLAGN NGC~4278 shows that both can account for the observed X-ray and $\gamma$-ray emission up to the TeV range in the quasi-quiet (low TeV flux) and active (high TeV flux) states, either via a moderately relativistic jet or a sub-relativistic wind. Parameter scans indicate that an evolving radiation zone,  coupled with enhanced accretion rates and/or particle injection luminosities, can drive the transition from quasi-quiet to active states. Importantly, the leptonic and lepto-hadronic origins can be tested and distinguished through observations in the MeV band, as well as across the $\gamma$-ray and very-high-energy (VHE) ranges up to $\sim100$ TeV$ - $ 1 PeV.

For the wind scenario, the TeV observations suggest that approximately $2.6-12\%$ of the accreted energy is converted into accelerated protons, making TeV-emitting LLAGNs a plausible contributor to the PeV diffuse neutrino background, considering that only a small fraction of an LLAGN's lifetime is spent in a TeV-emitting phase (TeV duty cycle $\Delta T_{\rm TeV}/T \sim 10^{-2}-10^{-3}$, based on LHAASO observations). 

Future multiwavelength observations of NGC 4278-like LLAGNs, along with the discovery of additional VHE-emitting LLAGNs, will allow more robust model selection and tighter constraints on baryon-loading efficiency, local source density, and TeV duty cycles. Together with this study, these efforts will enhance our understanding of LLAGNs as potential sources of the diffuse neutrino background.

\section*{Acknowledgements}
We thank Federico Testagrossa, Pavlo Plotko, and Marc Klinger for their assistance in developing the parameter scan tools. {We also thank the anonymous referee for helpful and constructive comments.} The work is partially funded by IISN project No. 4.4501.18 (C.Y.). R.-Y. L. acknowledges the support by National Natural Science Foundation of China under grants No.~12393852 and 12330006, and Basic Research Program of Jiangsu under grant No.~BK20250059. 





\appendix

\section{SED $\chi^2$-fitting and results}\label{app:fitting}
\begin{figure*}\centering
    \includegraphics[width=0.49\textwidth]{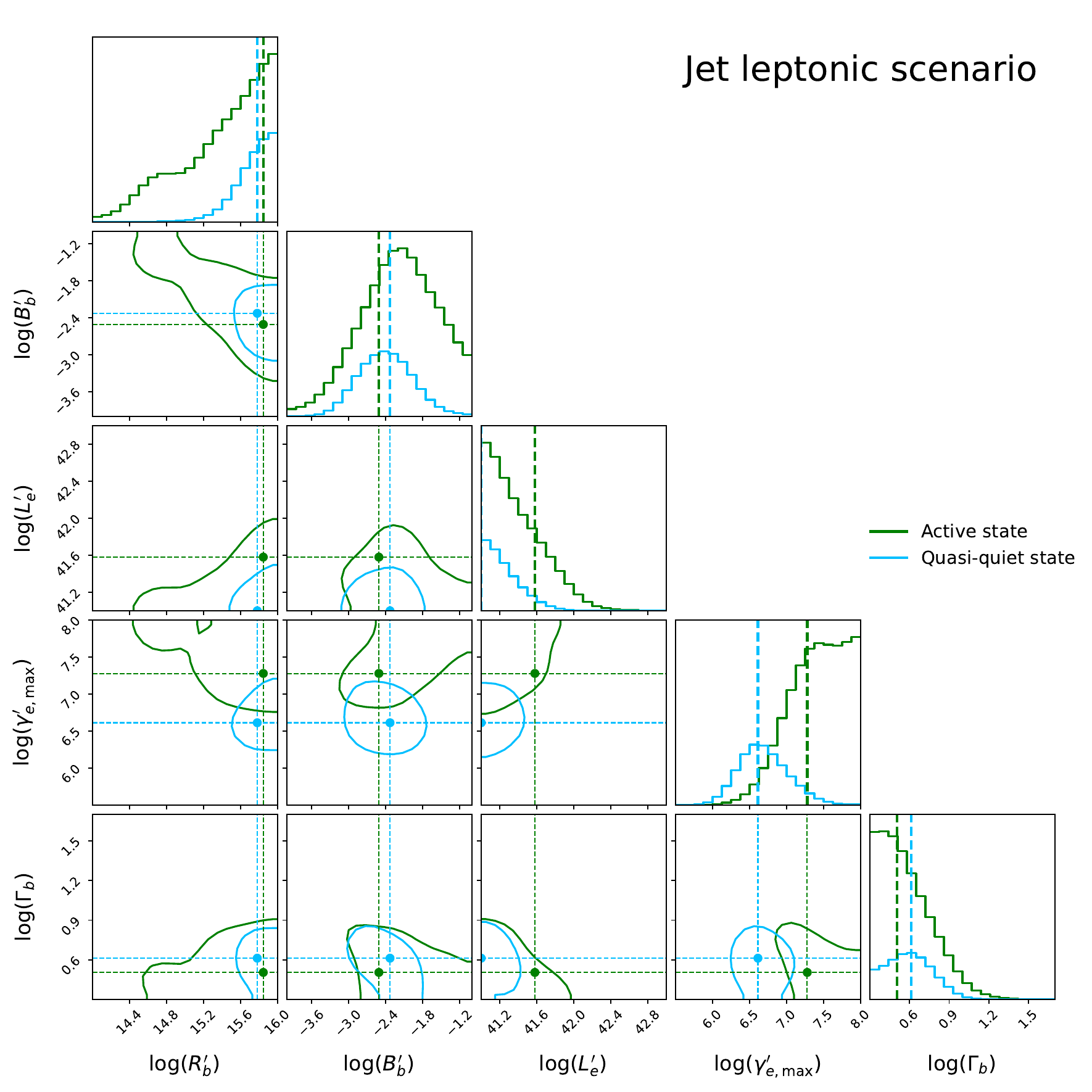}
        \includegraphics[width=0.49\textwidth]{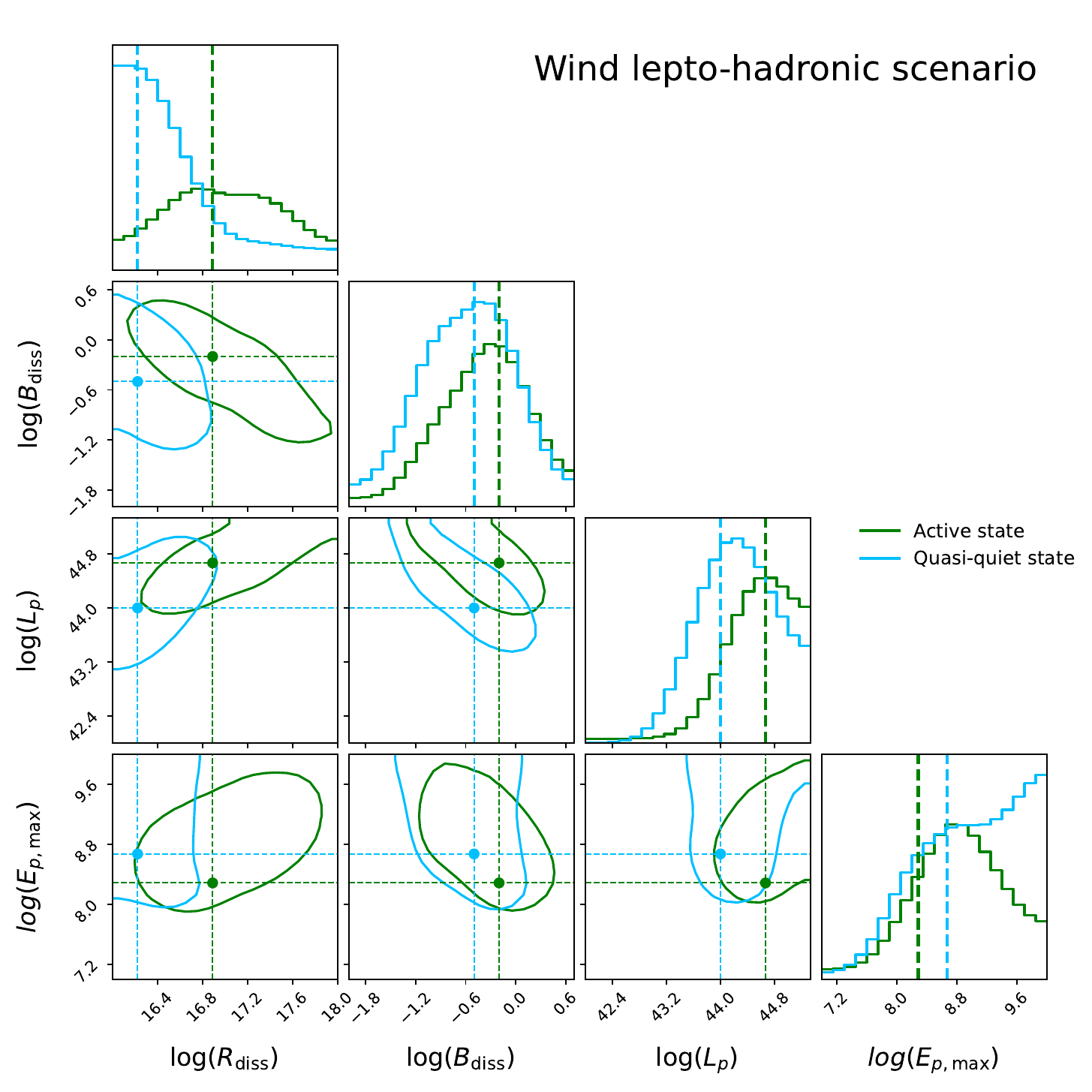}
    \caption{Corner plots illustrating the posterior parameter distributions for the jet leptonic (left panel) and wind lepto-hadronic (right panel) scenarios in the active (green) and quasi-quiet (blue) states. The contours indicate the $1\sigma$ uncertainties, while the best-fit parameters obtained by minimizing $\chi^2$ are marked by dots.}

    \label{fig:corner_plots}

\end{figure*}

We perform scans over the key parameters to fit the X-ray and $\gamma$-ray data. The free parameters and their ranges are listed in Table~\ref{tab:ranges}. To evaluate the goodness of fit, we construct the $\chi^2$ for each parameter set, incorporating both measurements and upper limits:\begin{linenomath}
\begin{equation}
\begin{split}
\chi^2 &= \sum_i \frac{[X_{\rm obs}(E_{\gamma, i}) - X_{\rm model}(E_{\gamma,i})]^2}{\sigma_i^2}  - \sum_i \ln \Phi\left[\frac{X_{\rm ul}(E_{\gamma, i}) - X_{\rm model}(E_{\gamma,i})}{\sigma_{{\rm ul},i}}\right],
\end{split}
\label{eq:app_chi2}
\end{equation}\end{linenomath}
where $X_{\rm obs}(E_{\gamma, i})$ is the measured flux at energy $E_{\gamma, i}$ with error $\sigma_i$, $X_{\rm ul}(E_{\gamma, i})$ is the upper limit, and $X_{\rm model}(E_{\gamma,i})$ is the flux predicted by the model. To incorporate the upper limits, we adopt $\sigma_{{\rm ul},i} = 0.1 X_{\rm ul}(E_{\gamma,i})$ and use the cumulative distribution function $\Phi$ of the standard normal distribution:\begin{linenomath}
\begin{equation}
\Phi(x) = \frac{1}{\sqrt{2\pi}} \int_{-\infty}^x e^{-t^2/2} dt.
\end{equation}\end{linenomath}
For each model, the best-fit parameters are obtained by minimizing $\chi^2$. The left and right panels of Figure~\ref{fig:corner_plots} show the posterior distributions of the parameters for the jet and wind scenarios, respectively, derived from the likelihood function $\mathcal L\propto \exp(-\chi^2/2)$. In each panel, the $1\sigma$ error contours and the marginalized 1D distributions are shown for the active (green) and the quasi-quiet (blue) states.

\begin{table*}
	\caption{Ranges of free parameters for the jet and wind scenarios.}
	\label{tab:ranges}\centering

       {\bf Jet leptonic scenario}
    \\
    \begin{tabular}{llc} 
		\hline
		  Parameter & Description & Range ($\log_{10}$-scale) \\
		\hline
		$R_b'~\rm [cm]$ &  Blob radius & [14,~16] \\
        $B_b'~\rm[G]$ &  Magnetic field strength & [-4,~-1]\\
        $L_e'~\rm[erg~s^{-1}]$ & Electron luminosity &[41,~43]\\
        $\gamma_{e,\rm max}'$ & Maximum electron Lorentz factor & [5.5,~8]\\
        $\Gamma_b$ & Jet Lorentz factor & [0.1,~$\log_{10}(50)$]\\
        
		\hline
	\end{tabular}

\vspace{1em} 

 {\bf Wind lepto-hadronic scenario}\\   
	\begin{tabular}{llc} 
		\hline
		  Parameter & Description & Range ($\log_{10}$-scale) \\
		\hline
		$R_{\rm diss}~\rm [cm]$ &  Energy dissipation radius & [16,~18] \\
        $B_{\rm diss}~\rm[G]$ &  Magnetic field strength & [-2,~$\log_{10}(5)$]\\
        $L_p~\rm[erg~s^{-1}]$ & Proton luminosity &[42,~45]\\
        $E_{p,\rm max}~[\rm GeV]$ & Maximum proton energy & [7,~10]\\   
		\hline
	\end{tabular}

\end{table*}

\newpage
\bibliography{ref.bib}
\end{document}